\documentstyle[12pt]{article}
\newcommand{\gsim}{ \raisebox{-.5ex}{\mbox{$\,\stackrel{>}{\sim}\,$}} }
\newcommand{\lsim}{ \raisebox{-.5ex}{\mbox{$\,\stackrel{<}{\sim}$\,}} }
\begin{document}
\baselineskip 8mm

\begin{center}
{\large\bf BIG ENTROPY FLUCTUATIONS\newline
 IN STATISTICAL EQUILIBRIUM:\newline
     THE MACROSCOPIC KINETICS}\\[5mm]
                   Boris Chirikov and Oleg Zhirov\\
{\it Budker Institute of Nuclear Physics \\
        630090 Novosibirsk, Russia}\\[1mm]      
        chirikov @ inp.nsk.su\\
        zhirov @ inp.nsk.su\\[5mm]
\end{center}

\begin{abstract} 
Large entropy fluctuations in an equilibrium steady state 
of classical mechanics
were studied
in extensive numerical experiments on a simple 2--freedom strongly chaotic 
Hamiltonian model 
described by the modified Arnold cat map.
The rise and fall of a large separated fluctuation was shown to be described by
the (regular and stable) "macroscopic" kinetics both fast (ballistic) and
slow (diffusive). We abandoned a vague problem 
of "appropriate" initial conditions 
by observing (in a long run)
spontaneous birth and death of arbitrarily big fluctuations
for any initial state of our dynamical model.
Statistics of the infinite chain of fluctuations,
reminiscent to the Poincar\'e recurrences, was shown to be
Poissonian. A simple empirical relation for the mean period
between the fluctuations (Poincar\'e "cycle") has been found
and confirmed in numerical experiments.
A new representation of the entropy via the variance of only
a few trajectories ("particles") is proposed which greatly
facilitates the computation, being at the same time fairly
accurate for big fluctuations.
The relation of our results to a long standing debates over statistical
"irreversibility" and the "time arrow" is briefly discussed too.
\end{abstract}

\vspace{5mm}


\newpage


\section{Introduction:\newline
macroscopic vs. microscopic fluctuations}

\hspace*{\parindent} Fluctuations are inseparable part of 
the statistical laws.
This is well known since Boltzmann.
What is apparently less known are the peculiar properties of rare 
big fluctuations
(BF) as different from, and even opposite in a sense to, 
those of small stationary 
fluctuations. 
In this paper we consider the simplest type of chaotic dynamical systems,
namely a finite--freedom Hamiltonian system which admits the (stable)
{\it statistical equilibrium} (SE). This class of dynamical models is
still popular (since Boltzmann!)
in debates over the dynamical foundations of statistical mechanics 
(see, e.g., "Round Table on Irreversibility" in \cite{1}, and \cite{2}).

A fairly simple picture of BF in such systems 
is well understood by now, though not yet 
well known. To Boltzmann such a picture was the basis of his fluctuation
hypothesis for our Universe. Again, as is well understood by now such a
hypothesis is completely incompatible with the present structure of the
Universe as it would immediately imply the notorious "heat death" (see, e.g.,
\cite{3}). For this reason, one may even term such systems the {\it heat
death models}. Nevertheless, they can be and actually are widely used 
in the description
and study of local statistical processes in {\it thermodynamically} closed
systems. The latter term means the absence of any heat exchange with the
environment. Notice, however, that under conditions of the exponential 
instability of motion, typical for chaotic systems,
the only {\it dynamically} closed system would be the "whole Universe". 
Particularly, this excludes
the hypothetical "velocity reversal" also popular in debates
over "irreversibility" since Loschmidt 
(for discussion see, e.g., \cite{4}).

In any case, dynamical models with SE do not tell us the whole story of
either the Universe or even a typical macroscopic process therein.
The principal solution of this problem, unknown to Boltzmann, is quite
clear by now, namely, the "equilibriumfree" models are wanted.
Various classes of such models are intensively studied today.
Moreover, the celebrated cosmic microwave background tells us that 
our Universe {\it was born} already in the state of a heat death which, 
however, fortunately to us all became {\it unstable} 
due to the well--known Jeans gravitational instability
\cite{5}. This resulted in developing of a rich variety of collective
processes, or {\it synergetics}, the term recently introduced or, 
better to say, put in use by Haken \cite{6}.
The most important peculiarity of such a collective instability is in that
the total overall relaxation (to somewhere ?) with  ever increasing total 
entropy 
is accompanied by an also increasing phase space {\it inhomogeneity} 
of the system, particularly in temperature.
In other words, the whole system as well as its local parts become 
more and more
{\it nonequilibrium} to the extent of the birth of a {\it secondary} dynamics
which may be, and is sometime, as perfect as, for example, the celestial 
mechanics 
(for general discussion see, e.g., \cite{4,7,8}).

We stress that all these inhomogeneous nonequilibrium structures are not
BF like in SE but are a result of regular collective instability,
so that they are immediately formed under a certain condition.
Besides, they are typically {\it dissipative structures} in Prigogine's 
term \cite{9} due to exchange of energy and entropy with the
{\it infinite} environment. The latter is the most important feature of such
processes, and at the same time the main difficulty in studying the
dynamics of those
models both theoretically and in numerical experiments which are so much
simpler for SE systems.

In the latter case a BF consists of the two symmetric parts: 
the rise
of a fluctuation followed by its return, or relaxation,
back to SE (see Figs.1 and 2 below).
Both parts are described by the same kinetic (e.g., diffusion) equation,
the only difference being in the sign of time. This relates the
time--symmetric dynamical equations to the time--antisymmetric
{\it kinetic} (but not statistical!) equations. The principal difference
between the both, some times overlooked, is in that the kinetic equations
are widely understood as describing the relaxation only, i.e. 
{\it increase} of the entropy in a closed system, whereas in fact they do so
(at least in SE)
for the rise of BF as well, i.e. for the entropy {\it decrease}.
All this was qualitatively known already to Boltzmann \cite{10}.
The first simple example of symmetric BF
was considered by Schr\"odinger \cite{11}. Rigorous mathematical theorem
for the diffusive (slow) kinetics was proved by Kolmogorov in 1937
in the paper entitled "Zur Umkehrbarkeit der statistischen Naturgesetze"
("Concerning reversibility of statistical laws in nature")
\cite{12} (see also
\cite{13}). Regrettably, the principal Kolmogorov theorem still remains
unknown to both the participants of hot debates over "irreversibility" 
as well as the physicists actually studying
such BF (see, e.g., \cite{14}).

By now, there exists the well developed ergodic theory of dynamical
systems (see, e.g., \cite{15}). Particularly, it proves that the relaxation
(correlation decay, or mixing) proceeds eventually in both directions of time
for almost any initial conditions of a chaotic dynamical system.
However, the relaxation does not need to be always monotonic
which simply means
a BF on the way, depending on the initial conditions.
To get rid of such an apparently confusing (to many) "freedom" we take
a different approach to the problem: instead discussing the "true" initial
conditions and/or a "necessary" restriction of those we start our numerical
experiments at {\it arbitrary} initial conditions
(most likely corresponding to SE), and do {\it observe} what the dynamics 
and statistics of BF are like. Of course, such an approach is based on
a fundamental hypothesis that all the statistical laws are contained in,
and can be principally derived from, the underlying fundamental (Hamiltonian)
dynamics. To the best of our knowledge, there is as yet no contradiction
to this principal hypothesis. Notice, however, that such an approach 
can be directly
applied to the fluctuations in finite systems
with statistical
equilibrium only (for discussion see \cite{4} and \cite{16}).
In such, and only such, systems infinitely many BF grow up spontaneously 
{\it independent} of the initial conditions of the motion. This is similar
to the well--known Poincar\'e recurrences (for farther discussion see
Section 4 below). 

In spite of the essential restrictions the simple
SE models allow us to better understand the mechanism and role of BF in the
statistical physics. Besides removal of a vague problem of the initial
conditions such models help a lot in clarifying the relation between
macroscopic and microscopic description of chaotic systems.
Particularly, spontaneous rise of a BF out of SE is a {\it macroscopic} event 
as well as its subsequent relaxation back to SE, even in a few--freedom
system. Like other macroscopic processes the BF
are not only perfectly regular by themselves but also
surprisingly stable against any perturbations, both regular and chaotic.
Moreover, the perturbations
do not need to be small. At first glance, it looks very strange
in a chaotic, highly unstable, dynamics. The resolution of
this apparent paradox is in that the dynamical instability of motion
does affect the BF instant of time only. 
As to the BF evolution, it is determined by the kinetics whatever its 
mechanism,
from purely dynamical one, like in model (2.2) we will use in this paper, 
to a completely noisy (stochastic) one.
As a matter of fact, the fundamental Kolmogorov theorem \cite{12} is related
just to the latter case but remains valid in a much more general situation.
Surprising stability of BF is similar to the less known
conception of
robustness for the Anosov (strongly chaotic) systems \cite{17} whose
trajectories get only slightly deformed under a small perturbation 
(for discussion see \cite{4}).

In the present paper we consider a particular type of BF which is characterized
by a large concentration of "particles" in a small 
phase space domain of a dynamical system. In other
words, "our" fluctuations are localized in phase space and
separated in time. A more accurate definition of these
fluctuations will be given below in Section 3 (see Eq.(3.6)). 
The same fluctuations 
in a stochastic
model (with noise) were studied in detail in \cite{14}. 
There exist, of course, many other fluctuations 
with their own peculiarities (see,  e.g., \cite{18}).
The primary goal of our studies was the macroscopic kinetics
of big fluctuations on the background of small stationary
microscopic fluctuations. A brief outline of our results
was presented in \cite{16}.

\section{A Hamiltonian model:\\
most simple but strongly chaotic}

The systems with SE allow for
very simple models in both the theoretical analysis as well as numerical
experiments which the latter are even more important for us.
In the present paper we will use one of the most simple and popular model 
specified
by the so--called Arnold cat map (see \cite{19,20}):
$$ 
   \begin{array}{ll}
   \overline{p}\,=\,p\,+\,x & mod\ 1 \\
   \overline{x}\,=\,x\,+\,\overline{p} & mod\ 1 
   \end{array} 
   \eqno (2.1)
$$
which is a linear canonical map on a unit torus. It has no parameters, and is
chaotic and even ergodic. The rate of the local exponential instability,
the Lyapunov exponent $\lambda =\ln{(3/2+\sqrt{5}/2)}=0.96$, implies
a fast (ballistic) kinetics with relaxation time $t_r\sim 1/\lambda\approx 1$.
Throughout the paper $t$ denotes the time in map's iterations.

A minor modification of this map:
$$ 
   \begin{array}{ll}
   \overline{p}\,=\,p\,+\,x\,-\,1/2 & mod\ C \\
   \overline{x}\,=\,x\,+\,\overline{p}\,-\,C/2 & mod\ 1 
   \end{array} 
   \eqno (2.2)
$$
where $C$ is a circumference of the phase space torus allows us
to study both a fast (exponential) ballistic kinetics ( for $C=1$) 
as well as the slow
(diffusive) relaxation in $p$ (for $C\gg 1$) with characteristic time 
$t_p\sim C^2/4D_p\gg 1$ where $D_p=1/12$ is 
the diffusion rate in $p$. 
In contrast to slow diffusion
in $p$, the relaxation time in $x$ does not depend on $C$ ($t_r\sim 1$)
so that subsequent values of $x$ are practically uncorrelated.
The map (2.2) has the (unstable) fixed point at
$x=x_0=1/2\,,\ p=p_0=C/2$.

A convenient characteristic of BF size is rms volume (area) in 2D phase space 
($x,p$)
$$
\sigma (t)=\sigma_{p}(t)\cdot \sigma_{x}(t) \eqno (2.3)
$$
occupied by a group of $N$ trajectories (particles).
In ergodic motion at equilibrium $\sigma =\sigma_0=C/12$.
Due to a severe restriction to small $N\lsim 10$ in the numerical experiments
(see below) we have to use simple (average) characteristics only like (2.3).
On the other hand, these are just the macroscopic variables we are
interested in.

Below we also restrict ourselves to a particular case of BF with the fixed
prescribed position in the phase space:
$$
   x_{fl}\,=\,x_0\,=\,{1\over 2}\,, \qquad  p_{fl}\,=\,p_0\,=\,{C\over 2}
   \eqno (2.4)
$$
Then, the variance of phase space size 
$v=\sigma^2=\sigma_{p}^2\cdot \sigma_{x}^2$ is determined by the relation
$$
   \sigma_{p}^2\,=\,\langle p^2\rangle\,-\,p_0^2\,, \qquad
   \sigma_{x}^2\,=\,\langle x^2\rangle\,-\,x_0^2 \eqno (2.5)
$$
where brackets $\langle ...\rangle $ denote the averaging over
$N$ trajectories.
In ergodic motion at equilibrium $v=v_{SE}=C^2/12^2$. In what follows we will
use the dimensionless measure $\tilde{v}=v/v_{SE}\to v$,
and omit tilde. In diffusive approximation of the kinetic
equation the variable $v(t)$ is especially convenient as it
is varying in proportion to time. Moreover, $v\to v_p$ in
this case because of a quick relaxation $v_x\to 1$ in $x$.

In all advantages of variable $v$ its relation to the fundamental conception
of entropy is highly desirable. The standard definition of the entropy,
which can be traced back to Boltzmann, reads:
$$
   S\,=\,-\,\langle\,\ln{f(x,p)}\,\rangle\,+S_0  \eqno (2.6)
$$
where $f(x,p)$ is a coarse--grained distribution function, or the phase--space
density, and $S_0$ an arbitrary constant to be fixed later. 
Notice that the distribution calculated from any {\it finite}
number of trajectories is always 
a coarse--grained one.
However, the direct application of Eq.(2.6) requires too many trajectories,
especially for BF of a small size. Nevertheless, just in the latter case,
which is the main problem under consideration, we have found
a simple approximate relation
$$
   S(t)\,\approx\,{1\over 2}\ln{v(t)} \eqno (2.7)
$$
which, at least, gives a rough estimate for the entropy evolution \cite{16}.
Moreover, if the distribution is Gaussian 
$$
   f(x,p)\,\to\,f(p)\,=\,\frac{\exp{(-(p-p_0)^2/2v)}}{\sqrt{2\pi v}} 
   \eqno (2.8)
$$
the estimate (2.7) becomes exact as it is directly followed
from the definition of the entropy (2.6). 
Comparison of both relations for the entropy will be considered
by the end of next Section for a typical BF.

A great advantage of (2.7) is
in that the computation of $S$ does not require very many trajectories
as does the distribution function. In fact, even a single trajectory is
sufficient as Fig.1 in \cite{16} and below demonstrate !

A finite number of trajectories used for calculating the variance $v$
is a sort of the coarse--grained distribution,
as required in relation (2.6), but with a free bin size which can be 
arbitrarily small.

Now we may turn to the numerical experiments.

\section{Macroscopic kinetics:\\
 complete, regular, and stable}

In this Section we consider the regular BF kinetics.
The data were obtained from simultaneous running of $N$ trajectories for 
very
long time in order to collect sufficiently many BF for reliable separation of
the regular part of BF, or the kinetic subdynamics in Balescu's term
(see \cite{21} and references therein), from the stationary fluctuations.
The separation was done by the plain
averaging of individual $v_i$ values ($i=1,...,n$) over all $n$ BF collected 
in a run. 

The size of BF chosen for further analysis was 
fixed by the condition that current
$$
   v(t) < v_b \eqno (3.1)
$$
at some time instant $t\approx t_i$, the moment of BF. Here a prescribed
$v_b$ is the main input parameter of the run.
This condition
determines, in fact, the border of the whole fluctuation
domain (FD): $0 < v < v_b$.

The event of entering FD is the macroscopic "cause" of BF whose obvious
"effect" {\it will be subsequent} relaxation to the equilibrium.
However, and this is the main point of our study, the second "effect"
of the same "cause" {\it was preceding} rise of BF in apparent contradiction
with the "causality principle" (for discussion see \cite{16} and Section 4
below). In any event, the second effect requires the permanent memory of
trajectory segments within some time window $w$ which is another important 
input parameter of the run. 

The exact procedure of data processing during the run was as follows.
Starting from arbitrary (random) initial conditions the selection rule (3.1)
is checked at each iteration. Suppose, it is satisfied at some instance
$t_{in}$ when the bundle of trajectories enters FD. 
In the first approximation we could consider it as the fluctuation
maximum (or variance minimum):
$t_i=t_{in}$ where sub $i$ is the number of current fluctuation
in a run. However, such a simple procedure would cause asymmetry with
respect to $t=t_i$. A better choice were the rule: $t_i=(t_{in}+t_{out})/2$
where $t_{out}$ is the exit moment of time from FD.
Instead, we have accepted the following, a more complicated  procedure which, as we hope, better restores the true BF symmetry.
Starting from the moment $t_{in}$ we search for
the minimum of $v(t)$ inside a rather large interval 
$t_{in}< t < t_{in}+w$.
If a minimum is found at some $t=t_{min}$, we check that it is the minimum inside
the next interval $t_{min}< t < t_{min}+ w$ too. If so, we 
identify this minimum with
the BF top, and set: $t_i=t_{min}$, otherwise we put $t_{min}$ to the time of
better minimum and repeat the last step again. Obviuosly, the parameter $w$
should be small compared to $\langle P\rangle$, the mean period of BF, but 
sufficiently long for the trajectory to leave the FD (3.1).
Typically,
$w\gsim C^2$, the total diffusion time, was chosen.
After fixing the current $t_i$ value the computation within
the interval $t_i<t<t_i+w$ had been completed, and only then
the search for the next BF was continued.

As was already mentioned above, there are two rather simple limiting cases 
of generally very complicated kinetics, namely the fast (ballistic) and
slow (diffusive) ones. An example of both in one run 
for $N=1$ (!) is presented in Fig.1 for two fluctuations of
different size.
In this case the general condition (3.1)
was checked separately for $p$ and $x$:
$$
   v_p(t)\,<\,v_{pb}  \quad and \quad   v_x(t)\,<\,v_{xb} \eqno (3.2)
$$
with $v_{pb}=v_{xb}\sim 10^{-5}$ and $v_b=v_{pb}\cdot v_{xb}\sim 10^{-10}$. 

The fast part of kinetics is approximately described as
$$
   v(\tau )\,\approx\,v(0)\cdot\exp{(4\lambda\tau )} \eqno (3.3)
$$
where $\tau =t-t_i,\ \lambda$ the Lyapunov exponent (Section 2), 
and where $v(0)\sim 10^{-13}$ is the minimal
variance averaged over all $n$ fluctuations observed in the run.
Notice that the latter value is considerably smaller than the border
$v_b\sim 10^{-10}$. This is because of penetration of trajectories into FD.
Interestingly, the ratio
$v_b/v(0)=2000$ is the same for both runs in Fig.1

Surprisingly sharp crossover to diffusive kinetics, clearly seen in Fig.1, 
is related to
the dynamical scale of diffusion which corresponds to a certain
size $v_d$ of the increasing variance at which the exponential growth stops.
Roughly, it occurs at time $\tau =\tau_d$ when already 
$|x-x_0|\sim |p-p_0|\sim 1/2$, 
whence $v_{xd}\sim 12/4=3$ and $v_{pd}\sim 3/C^2$.
Hence, we can characterize the dynamical
scale as
$$
   v(\tau_d)\,=\,v_d\,=\,F_d\cdot v_{pd}\cdot v_{xd}\,=\,{9F_d\over C^2}\,, 
   \qquad \tau_d = {\ln{(v_d/v(0))}
\over 4\lambda} 
\eqno (3.4)
$$
where $F_d$ is an empirical factor, and $\tau_d$ is found
from Eq.(3.3).
From data in Fig.1 the dynamical scale 
$v_d\approx 0.015$ independent of $v_b$ which gives the empirical factor 
$F_d\approx 1/3$.

In the diffusion region ($v>v_d$) the initial kinetics is described by a simple
relation for the free diffusion (see Section 2):
$$
   v(\tau )\,\approx{\tau\pm\tau_d\over C^2}\,+\,v_d\,, 
   \qquad \tau_d\,<\,\tau\,\ll\,C^2
   \eqno (3.5)
$$
which is also shown in Fig.1.
It includes two corrections, $\tau_d$ and $v_d$, due to the
exponential ballistic kinetics. The first one with
opposite signs for two symmetric parts of the fluctuation
takes account for the "lost" time after (or prior to) 
anti--diffusion (diffusion) while the second correction
describes a finite fluctuation size at the crossover
from (to) the diffusion. The mean empirical $\tau_d=7$, used
in Fig.1, is close to the value $\tau_d=6.5$ found from
Eq.(3.4) with another empirical quantity $v_d=0.015$.

The large ratio
$$
   B\,=\,{\langle P\rangle\over C^2}\,\gg\,1 \eqno (3.6)
$$
of the mean fluctuation period $\langle P\rangle$ to a characteristic time of 
diffusion relaxation (see Eq.(3.5)) is the definition of {\it big}
fluctuation. It guaranties the separation of successive fluctuations in time.

Now we turn to the main subject of our studies, the purely
diffusive kinetics of big fluctuations. To this end we, first,
get rid of $x$--statistics excluding $v_x$ from the
selection condition (3.1) which now reads:
$$
   v(t)\,=\,v_p\,<\,v_{pb}\,=\,v_b \eqno (3.7)
$$
Besides, the variance $v_b$ must now exceed the new dynamical
border:
$$
   v_b\,>\,v_d\,=\,v_{pd}\,\approx\,f_p\cdot{12\over C^2}
   \eqno (3.8)
$$
with some empirical factor $f_p\approx 1$ (see Eq.(3.4), and
discussion below).

A typical example of diffusive BF is shown in Fig.2.
Both the regular macroscopic kinetics of 
anti--diffusion/diffusion as well as irregular fluctuations around
are clearly seen. Notice that their
size is rapidly
decreasing toward the BF maximum. One may even get the impression that
the motion becomes regular over there, hence the term "optimal
fluctuational path" \cite{14}. In fact, the motion remains diffusive down
to the dynamical scale $v\sim v_d$ (3.8).

Even though a separate BF by itself is fairly regular,
the time instance of its spontaneous appearance $t_i$, and
hence the individual period $P$, are 
random in the chaotic system. Due to statistical independence of BF under
condition (3.6) the expected distribution in $P$ is
Poissonian (Fig.3):
$$
   f(P)\,=\,{\exp{(-P/\langle P\rangle )}\over \langle P\rangle } \eqno (3.9)
$$

The principal characteristic of the period statistics, $\langle P\rangle$,
can be estimated as follows. From the ergodicity of motion
in the $N$--dimensional momentum space the ratio
$$
   \Phi\,=\,{T_s\over t_f}\,=\,{\langle T_s\rangle\over
   \langle P\rangle }\,=\,{{\cal{P}}_{fl}\over {\cal{P}}_{eq}}
   \eqno (3.10)  
$$
This is exact relation (in the limit $t_{run}\to\infty$) where
$T_s$ is the total sojourn time of trajectories within FD
(under condition $v(t)<v_b$) during
the whole run time $t_{run}$, and $\langle T_s\rangle$ is the same per
fluctuation. Both ratios are equal to the ratio of
$N$--dimensional momentum volume 
$\cal{P}$ of the fluctuation at $\tau =0$ to that in
equilibrium. The ratio $\Phi$ was also measured during the run.
Hence
$$
   \langle P\rangle\,=\,{\langle T_s\rangle\over \Phi} \eqno (3.11)
$$

The next more difficult step is evaluation of $T_s=2T_{ex}$ from the
diffusion equation where $T_{ex}$ is the exit (or entrance due to symmetry)
time from (or to) FD.
A simple crude estimate is: $T_{ex}\sim v_b/D_p=v_bC^2$ (Section 2).
However, the first numerical experiments have already revealed that the actual
exit time is much shorter, roughly by a factor of $1/N^2$.
A plausible explanation is in that the most of distribution inside FD
is concentrated in a relatively narrow layer at the surface
of $N$--dimensional sphere determined by the selection condition 
$v(t)<v_b$ (3.7). Then, the relative width of the layer $\sim 1/N$ implies the
observed factor $\sim 1/N^2$.
Father, the ratio
$$
   \Phi (v_b,N)\,=\,v_b^{N/2}\cdot \phi (N) \eqno (3.12)
$$
where the geometrical function
$$
  \phi (N)\,\approx\,\left({\pi e\over 6}\right )^{N/2}
  {(1-1/6N)\over \sqrt{\pi N}} \eqno (3.13)
$$
admits a fairly accurate approximation down to $N=1$
(see Fig.4).

Collecting all the formulae above, we arrive at our final
empirical relation
$$
   \langle P\rangle\,\approx\,{F\over \Phi}\cdot{2v_bAC^2\over N^2}\,\approx\,F\cdot{2AC^2\over N^2}\cdot
   {v_b^{1-N/2}\over \phi(N)} 
   \eqno (3.14)
$$
with two fitting factors, $A$ for the layer width, and
$F$ for all other approximations above.
Both factors cannot be united in one because the former
enters a new expression for the dynamical scale which naturally 
generalizes Eq.(3.8). Together with inequality (3.6) for big fluctuation
the new dynamical scale was using for selection of purely
diffusive BF which are described by Eq.(3.14).
The corresponding inequality reads (cf. Eq.(3.8)):
$$
   v_b\,>\,v_d\,, \qquad v_d\cdot{A\over N^2}\,\approx\,
   f_p\cdot{12\over C^2}
\eqno (3.15) 
$$
which means that even a small part ($A/N^2<1$)
of FD must exceed the dynamical scale.

Optimization of all empirical parameters was done as follows.
The values of two factors, $B$ in Eq.(3.6) and $f_p$ in (3.15),
are not crucial that is the larger they are the better for selection of purely diffusive BF. However, this reduces the
amount of empirical data available. A compromise was found
at $B=7$ and $f_p=1$ which leaves 36 runs of 61 done, and
34429 of overall 75053 BF computed with $N=2 - 10$ for
comparison to Eq.(3.14). This was executed in the following
way. For each selected run with parameters $N,\ C,\ v_b$, and
computed $\langle P\rangle$ and $\Phi$ the empirical factor $F$ 
supposed to be a constant was calculated from the first
Eq.(3.14). The value of parameter $A$ was chosen by
minimizing the relative standard deviation down to
$\Delta F/\langle F\rangle =0.17$.
The result, for a given set of
data, was $A\approx 6$. The final dependence $F(N)$ is shown
in Fig.5 where the bars are the statistical errors
$F/\sqrt{n}$ for each run.

Coming to analysis of our main theoretical result, the
second Eq.(3.14), we first remark that
it does not describe at all
a single trajectory ($N=1$). This is because we excluded
$v_{xb}$ from the selection condition (3.7) (cf. Eq.(3.2)),
and thus reduced the dimension of phase space to the minimal
value, the unity. In this case, a single trajectory 
repeatedly crosses FD with a period $P\sim C^2$, the whole
diffusion time around the phase space torus, independent of FD size. More formally, it
follows also from Eq.(3.14) since condition (3.6) cannot be
satisfied for small $v_b$.

In case of two trajectories ($N=2$) the period does not
depend on $v_b$, and for data in Fig.5 the ratio
$\langle P\rangle /C^2\approx 8.7$. Due to fluctuations
the actual values of this ratio are in the interval
$7.4 - 11.0$, still not too big for a Big Fluctuation.
Apparently, this leads to a relatively large scattering of
points with $N=2$ which also persists for $N=3$ too.

The main, exponential in $N$, dependence in Eq.(3.14) 
is readily derived from a graphic picture of $N$ statistically
independent particles gathering together inside a small domain
with probability $\sim 1/P\sim v_b^{N/2}$. 
Such estimates are
known for the Poincar\'e recurrences since Boltzmann \cite{10}.
The estimate is especially vivid in geometrical picture of
$N$--dimensional sphere of radius $\sqrt{v_b}$ considered above.
Our empirical relation (3.14) considerably improves the simple
estimate including
a more weak power--law dependence which is evident in Fig.5.

In our studies described above we fixed the position of BF
in the phase space, Eq.(2.4). If we lift this restriction,
the probability of BF would increase by a factor of 
$v_b^{-1/2}$, or by decrease of $N$ by one ($N\to N-1$)
because now only $N-1$ trajectories remain independent.
With the latter change all the above relations would
presumably still hold true.

Our main relation (3.14) describes the diffusive kinetics for
$v_b>v_d$, Eq.(3.15), when a big fluctuation is not too big.
In the opposite case $v_b\ll v_d$ of a very big fluctuation,
like in Fig.1, the dependence $\langle P(v_b)\rangle $ becomes
much simpler\\ (see Eqs.(3.11) - (3.13), and \cite{16}):
$$
   \langle P(v_b)\rangle\,=\,{\langle T_s\rangle\over \Phi}\, 
   \approx\,{2\over v_b^{N/2}\phi (N)}\,\approx\,
   2v_b^{-N/2}
   \eqno (3.16)
$$
This is explained by a fast exponential kinetics near BF top
(Fig.1) which implies the most short exit time 
$T_{ex}\approx 1$ and, hence, $T_{s}\approx 2$.
Indeed, for both BF in Fig.1 the empirical value of the product
$\langle P\rangle\Phi = 1.98$.

In conclusion of this Section we show in Fig.6 the
macroscopic kinetics of BF entropy, both "exact" (2.6),
calculated on the partition of the whole interval\\
($0<p<C$) into $N_p=401$ bins, and that
in our approximation (2.7). Both entropies were calculated
for the same 5 trajectories in one run. A necessary statistics
for exact entropy was obtained at the expense of a large
number $n=4580$ of fluctuations in the run.
For comparison of both entropies we need, first,
to adjust the constant $S_0$ in Eq.(2.6). 
As is easily verified, the Gaussian distribution (2.8) leads
exactly to the relation (2.7) if the constant
$$
   S_0\,=\,-\,{1\over 2}\ln{(2\pi e)}\,\approx\,-1.4189\,
   \approx\,-\sqrt{2}
   \eqno (3.17)
$$
Approximation (2.7) holds on the most part of BF except a relatively
small domain near the equilibrium where the distribution in $p$
approaches the homogeneous one. The exact entropy 
(with constant (3.17)) in equilibrium is
$$
   S_{SE}\,=\,-\,{1\over 2}\ln{\left({\pi e\over 6}\right)}\,
   \approx\,-0.18 \eqno (3.18)
$$
instead of zero in approximation (2.7). The difference is
relatively small, the smaller the bigger is the fluctuation.
In the main part of BF our simple relation of the entropy
(2.7) reproduces the exact one (2.6) to a surprisingly good
accuracy which confirms that the distribution in $p$ is indeed
very
close to the Gaussian one (2.8) as expected.

\section{Conclusion: thermodynamic arrow ?}
In the present paper the results of extensive numerical experiments on
big entropy fluctuations (BF) in a statistical equilibrium (SE)
of classical
dynamical systems are presented,
and their peculiarities are analysed and discussed. 

All numerical experiments have been carried out on the basis of a very simple
model - the Arnold cat map (2.1) on a unit torus - with only  two minor, but important and helpful,
modifications:

(1) Expansion of the torus in $p$ direction (2.2) which allows for more
    impressive diffusive kinetics of BF out of the equilibrium
    (Fig.2), and 
    
(2) Inserting a special (unstable) fixed point for a better
demonstration of exponential ballistic kinetics as well (Fig.1).
Besides, this point was used as a fixed position of BF, thus
relating our studies of BF to another interesting and
important problem, the Poincar\'e recurrences (see Eq.(2.2)).

The most important distinction of our approach to the problem
was in that we have abandoned from the beginning a vague
question of the initial conditions, particularly 
a "necessary" restriction of those in statistical physics.
Instead, we started our numerical
experiments at {\it arbitrary} initial conditions
(most likely corresponding to SE), and did {\it observe} what the dynamics 
and statistics of BF were like. In other words, we studied
the {\it spontaneous} BF only.

What is also important, such a spontaneous rise of BF out of SE 
as well as its subsequent relaxation back to SE
can be considered as a statistical {\it macroscopic} event,
even in a few--freedom
system like (2.2). The term "macroscopic" refers here to average quantities as variance, entropy, mean period,
distribution function, and the like.

We consider a particular class of BF which we call the Boltzmann
fluctuations. They are obviously symmetric with respect
to time reversal (see Figs. 1, 2 and 6), so that at least in this case there is no physical reason 
at all for the conception
of the notorious '{\it time} arrow'. 
Nevertheless, a related conception, say, {\it thermodynamic} arrow, pointing
in the direction of the average increase of entropy, makes sense in spite
of the time symmetry \cite{16}.
The point is that the BF characteristic
relaxation time 
is determined by model's parameter $C$ only, and does not depend on the BF
itself. On the contrary, the expectation time for a given BF,
or the mean period between successive fluctuations, rapidly grows 
with BF size and with the number of trajectories (or freedoms),
Eq.(3.14). A large ratio of both 
$B\,=\,\langle P\rangle / C^2\,\gg\,1$ is our {\it definition}
of big fluctuation, Eq.(3.6).
A similar result was obtained recently in \cite{22} but the
authors missed the principal difference between the time arrow
and thermodynamic arrow.

A related notion of {\it causality} arrow, which by definition
points from an {\it independent} macroscopic cause to its effect, 
also makes some
physical sense (for discussion see \cite{16} and Section 3 above).
For Boltzmann's
BF considered in the present paper the directions of both arrows
do coincide {\it independent} of the direction of time.
The latter statement is the most important, in our opinion,
philosophical "moral" the principally well--known Boltzmann
fluctuations do teach us.

Even though we prefer the discussion and interpretation of our empirical results in terms of entropy ($S$), the most
fundamental conception in
statistical physics, we actually use in our studies
another, entropy--like, quantity, the variance $v(t)$ for a group of
$N$ trajectories, Eq.(2.5). One reason is technical: the
computation of $v$ is much simpler while that for $S(t)$ is
either very time--consuming in numerical experiments 
(for exact $S$ (2.6)) or approximate (2.7). Besides, for
diffusive kinetics, we are mainly interested in, the variance
is natural variable in which the BF picture is most simple
and comprehensible.

Originally, we planned to cover both sides of the BF
phenomenon,
the regular macroscopic kinetics as well as accompanying
microscopic fluctuations (noise) around. However, our
numerical experiments revealed a much more complicated structure
of the latter as an example in Fig.2 demonstrates.
The dependence $v(t)$ looks like a fractal curve on a variety
of time scales, from the minimal one $\sim 1$ iteration
up to $\sim C^2$ comparable to that of BF itself.
This interesting problem certainly requires and deserves
farther special studies.

In the present paper the fluctuations in classical mechanics 
are considered
only. Generally, the quantum fluctuations would be rather 
different. However, according to the Correspondence Principle, the
dynamics and statistics of a quantum system in quasiclassics are close
to the classical ones on the appropriate time scales of which 
the longest one corresponds just to the diffusive kinetics,
and provides the necessary
transition to the classical limit (for details see \cite{4,23}).
Curiously, the computer classical dynamics that is the simulation
of a classical dynamical system on digital computer is of a qualitatively 
similar character. This is because any quantity in computer representation
is discrete ("overquantized"). As a result the correspondence between
the classical continuous dynamics and its computer representation in
numerical experiments is generally
restricted to certain finite time scales
like in the quantum mechanics (see two first references \cite{23}).

Discreteness of the computer phase space leads to another peculiar phenomenon:
generally, the computer dynamics is irreversible due to the rounding--off
operation unless the special algorithm is used in numerical experiments.
Nevertheless, this does not affect the statistical properties of chaotic
computer dynamics. Particularly, the statistical laws in computer 
representation remain time--reversible in spite of (nondissipative) 
irreversibility of the
underlying dynamics.
This simple example demonstrates that, contrary to a common belief,
the statistical reversibility is a more general property than the dynamical
one.


\newpage   
 \begin{center}  Figure captions   \end{center}

\begin{itemize}

\item[Fig.1] 
Mixed kinetics for two big fluctuations of different size. Full/open
circles show the time dependence 
of mean variance
$\langle v(t-t_i)\rangle$ around BF maximum at $t=t_i$; upper horizontal straight
line is equilibrium while the lower line indicates the
empirical value of
dynamical scale $v_d=0.015$, Eq.(3.4) with parameter
$F_d\approx 1/3$.
Two oblique straight lines represent
the expected fast kinetics, Eq.(3.3), and two solid curves do so for the
initial diffusive kinetics, Eq.(3.5). 
Run parameters and results are respectively: $C=15,\ N=1,\ 
v_b=3.9\times 10^{-11}
\ /\ 6.25\times 10^{-10}\ (v_{xb}=v_{pb}),\ v(0)=1.96\times 10^{-14}\ /\ 
3.1\times 10^{-13},\ n=1971\ /\  4459,\ w=500$.
Average period between successive fluctuations 
$\langle P\rangle\approx1.4\times 10^7\ /\  3.5\times 10^6$ iterations.

\item[Fig.2]
Same as in Fig.1 for
a typical diffusive kinetics (anti--diffusion/diffusion):
solid curve shows the average over all 
$n=20259$ fluctuations in a run, and wiggle line is the same for $28$ 
first fluctuations.
Two oblique straight lines represent
the expected
initial diffusive kinetics, Eq.(3.5) with $\tau_d=0$ and
empirical $v_d^{(emp)}=0.045$ while the theory (3.15) gives
$v_d=0.02$.
Other run parameters/results: $C=50,\ N=5,\ v_b=0.0256,\ w=10^4,\ \langle P\rangle\approx 7.7\times 10^5\ /\ 
8.7\times 10^5$, and
$B=306\ /\ 348;\ \langle P\rangle/w\approx 77\ /\ 87$.

\item[Fig.3]
Histogram of integrated distribution (3.9) for data in Fig.2.
Each circle shows the number of periods $P_m >m\cdot
\Delta P,\ m=0,1,...$ integer;\ $P_0=n,\ 
\Delta P=1.5\times 10^5;\ P_{min}/w=1.0027;\ 
P_{max}/\langle P\rangle =12.63;\ \langle P\rangle = 765084$.
Straight line is expected distribution 
$n\cdot\exp{(-P/\langle P\rangle )}$.

\item[Fig.4] 
Comparison of directly measured ratio $\Phi_{emp}$, Eq.(3.10), with
theoretical approximation $\Phi_{th}$, Eq.(3.12) for
$N=1\ -\ 10$:
$\Phi_1=\Phi_{emp}/\Phi_{th}$; average over 71 runs
$\langle\Phi_1\rangle = 1.015\pm 0.11$ (the standard deviation);
the bars show statistical errors $1/\sqrt{n}$ for each run;
the total number of fluctuations in all runs is 127346.

\item[Fig.5]
Comparison of empirical data for 36 runs, 
selected from 61 runs computed for $N=2\ -\ 10$ by the two rules,
Eq.(3.6) with $B>7$ and Eq.(3.15) with $A=6$,
to theoretical relation (3.14) with the main fitting factor
$F_m,\ m=1,...,36$ (see text). Average
$\langle F\rangle =1.51\cdot (1\pm 0.17)$ (the standard deviation);
the bars show statistical errors $F_m/\sqrt{n}$ for each run;
the total number of fluctuations in 36 runs is 34429.

\item[Fig.6]
Macroscopic kinetics of BF entropy: lower line is "exact"
entropy, Eq.(2.6), to be compared with approximation (2.7),
middle line; the upper line is the same approximation 
for diffusion theory, Eq.(3.5) with $\tau_d=0$ and
empirical $v_d^{(emp)}=0.02$.
Run parameters/results: $C=50,\ N=5,\ v_b=0.01,\ w=10^4,\ n=4580,\ 
 \langle P\rangle\approx 3.3\times 10^6,\  
B=1314;\ \langle P\rangle/w\approx 329$.
The number of partition bins for calculating (2.6) $N_p=401$.

\end{itemize}



\newpage

\begin{thebibliography}{99}
\bibitem{1} Proc. 20th IUPAP Intern. Conference on Statistical Physics
            (Paris, 1998), Physica A {\bf 263}, 516 - 544 (1999).
\bibitem{2} J. Lebowitz, Physica A {\bf 194}, 1 (1993).
\bibitem{3} L.D. Landau and E.M. Lifshitz, 
             {\it Statistical Physics}, Part 1,
             Moskva, Nauka, 1995 (English translation of a previous Edition:
             Pergamon, Oxford, 1980).
\bibitem{4} B.V. Chirikov, Natural Laws and Human Prediction, in:
             {\it Law and Prediction in the Light of Chaos Research},
             ed. by P. Weingartner and G. Schurz, Springer, Berlin, 1996, p. 10;
             Open Systems \& Information Dynamics {\bf 4}, 241 (1997);
             chao--dyn/9705003;
             Wiss. Zs. Humboldt Univ. zu Berlin, 
             Ges.-Sprachw. R. {\bf 24}, 215 (1975).             
\bibitem{5} J. Jeans, Phil. Trans. Roy. Soc. A {\bf 199}, 1 (1929).
\bibitem{6} H. Haken, {\it Synergetics}, Springer, Berlin, 1978 
            (Russian translation:
             Moskva, Mir, 1980).
\bibitem{7} A. Turing, Phil. Trans. Roy. Soc. Lond. B {\bf 237}, 37 (1952);
             G. Nicolis and I. Prigogine, {\it Self--Organization in 
             Nonequilibrium Systems}, Wiley, New York, 1977.
\bibitem{8} A. Cottrell, Emergent properties of complex systems,
             in {\it The Encyclopedia of Ignorance}, ed. by
             R. Duncan and M. Weston-Smith, Pergamon, Oxford,  1977, p.129.
\bibitem{9} P. Glansdorf and I. Prigogine, {\it Thermodynamic theory of
             structure, stability, and fluctuations}, Wiley, New York, 1971
             (Russian translation: Moskva, Mir, 1972).             
\bibitem{10} L. Boltzmann, {\it Vorlesungen \"uber Gastheorie}, Leipzig, Barth,
             1896/98 (English translation: 
             {\it Lectures on gas theory}, Cambridge Univ. Press, 1964;
             Russian translation: Moskva, Gostexizdat, 1956).
\bibitem{11} E. Schr\"odinger, \"Uber die Umkehrung der Naturgesetze,
            Sitzungsber. Preuss. Akad. Wiss. (1931), 144.
\bibitem{12} A.N. Kolmogoroff, Math. Ann. {\bf 113}, 766 (1937);
             see also ibid. {\bf 112}, 155 (1936) (Russian translation:
             {\it Selected papers on probability theory and
             mathematical statistics}, ed. by Yu.V. Prokhorov, Moskva, Nauka,
             1986, pp. 197 and 173).
\bibitem{13} A.M. Yaglom, Dokl. Akad. Nauk SSSR {\bf 56}, 347 (1947);
            Mat. Sbornik {\bf 24}, 457 (1949).            
\bibitem{14} D.G. Luchinsky, P. McKlintock and M.I. Dykman,
             Rep. Prog. Phys. {\bf 61}, 889 (1998).                          
\bibitem{15} I.P. Kornfeld, S.V. Fomin and Ya.G. Sinai, {\it Ergodic Theory},
            Moskva, Nauka, 1980 (English translation:
            Springer, New York, 1982). 
\bibitem{16} B.V. Chirikov, Big Entropy Fluctuations in Nonequilibrium Steady
              State: A Simple Model with Gauss Heat Bath,
                            preprint Budker INP 00--54; nlin.CD/0006033;
                            JETP (in press).
\bibitem{17} D.V. Anosov, Dokl. Akad. Nauk SSSR {\bf 145}, 707 (1962).
\bibitem{18} L. Schulman, Phys. Rev. Lett. {\bf 83}, 5419 (1999);
             G. Casati, B.V. Chirikov and O.V. Zhirov, ibid. {\bf 85}, 896 
             (2000);
             D. Evans, D. Searles, E. Mittag, cond--mat/0008421;
             B.V. Chirikov and O.V. Zhirov, cond--mat/0009125.

\bibitem{19} V.I. Arnold and A. Avez, {\it Ergodic Problems of Classical 
            Mechanics}, Benjamin, New York, 1968 (Russian translation: RCD,
            Izhevsk, 1999).
\bibitem{20} A. Lichtenberg and M. Lieberman, {\it Regular and Chaotic 
             Dynamics},
            Springer, New York, 1992 (Russian translation of the 1st Edition:
            {\it Regular and Stochastic Motion}, Moskva, Mir, 1984).
\bibitem{21} R. Balescu, {\it Equilibrium and Nonequilibrium Statistical
             Mechanics}, Wiley, New York, 1975 (Russian translation:
             Moskva, Mir, 1978).
\bibitem{22} R. Metzler, W. Kinzel and I. Kanter,
On time's arrow in Ehrenfest models with reversible 
deterministic dynamics, cond--mat/0007382.
\bibitem{23} B.V. Chirikov, F.M. Izrailev and D.L. Shepelyansky,
             Sov. Sci. Rev. {\bf C2}, 209 (1981);
             B.V. Chirikov, Time-dependent quantum systems, Lectures in
             Les Houches Summer School on Chaos and Quantum Physics (1989),
             Elsevier, Amsterdam, 1991, p. 443;
             G. Casati and B.V. Chirikov, in: {\it Quantum Chaos:
             Between Order and Didorder}, ed. by G. Casati and B.V. Chirikov,
             Cambridge Univ. Press, Cambridge, 1995, p. 3; Physica D {\bf 86}, 220 (1995);
             B.V. Chirikov, Pseudochaos in statistical physics,
             Proc. Intern. Conference on Nonlinear Dynamics, Chaotic and
             Complex Systems (Zakopane, 1995), ed. by E. Infeld, R. Zelazny
             and A. Galkowski, Cambridge Univ. Press, Cambridge, 1997, p.149;
             B.V. Chirikov and F. Vivaldi, Physica D {\bf 129}, 223 (1999).

\end{thebibliography}
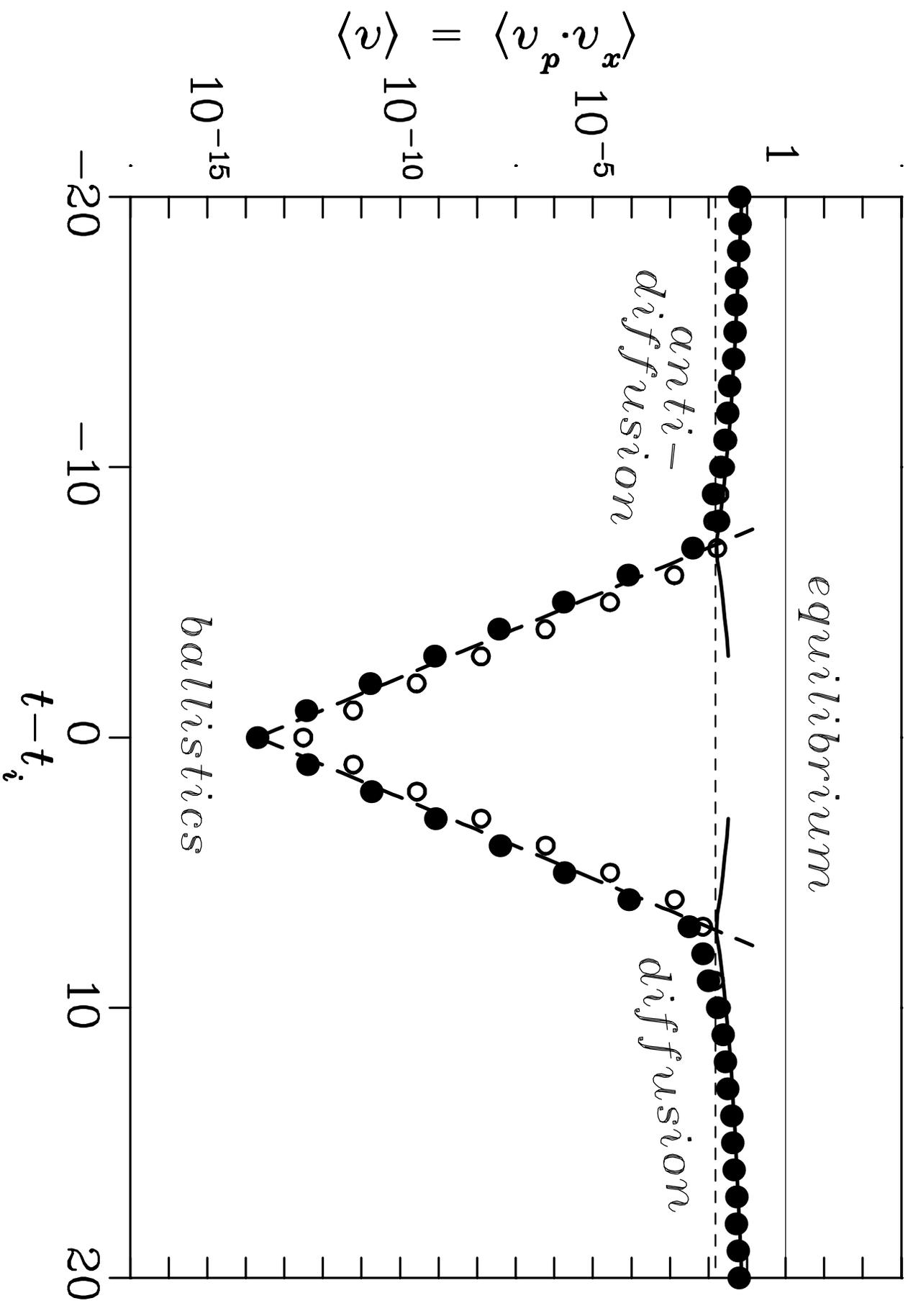
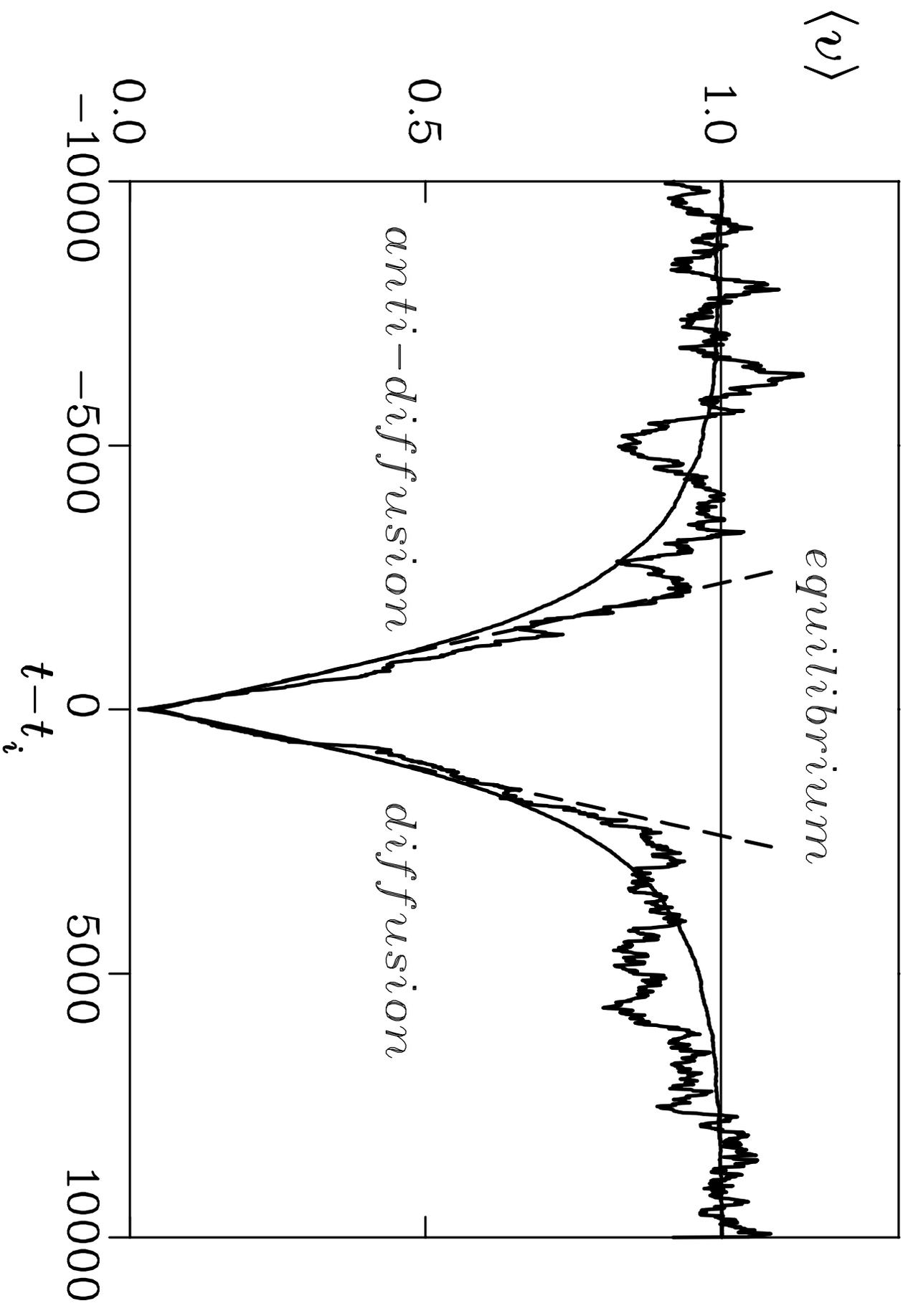
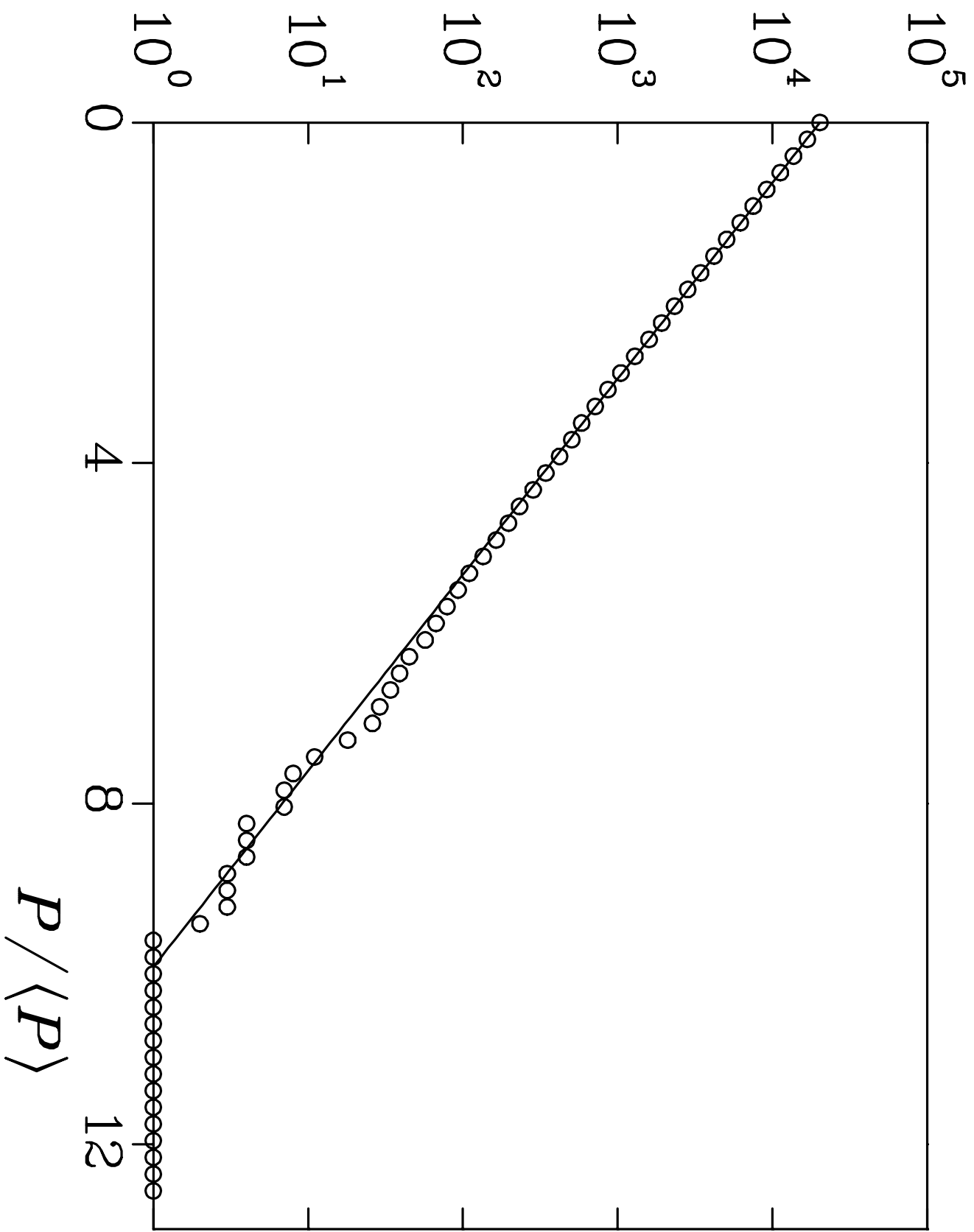
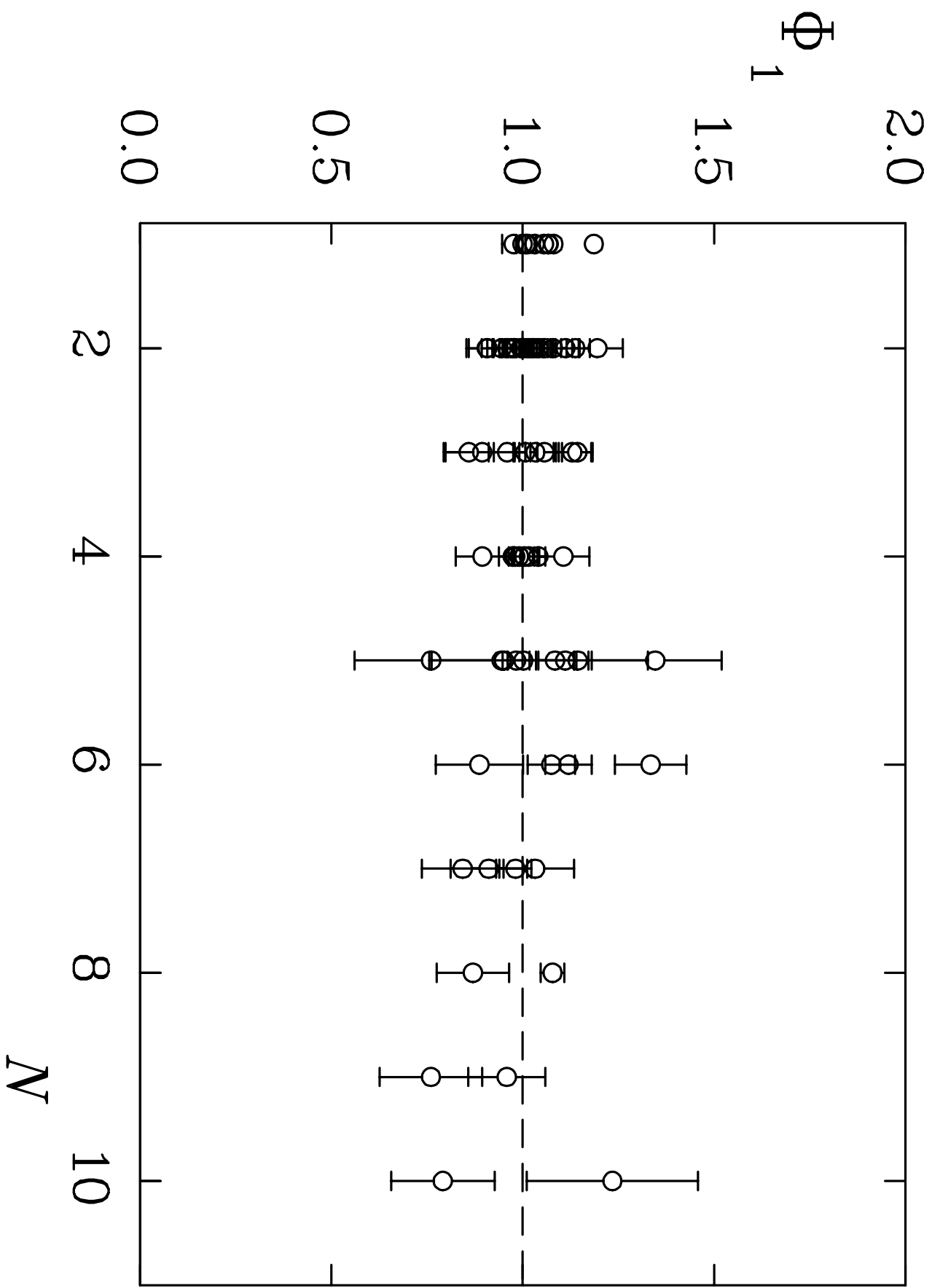
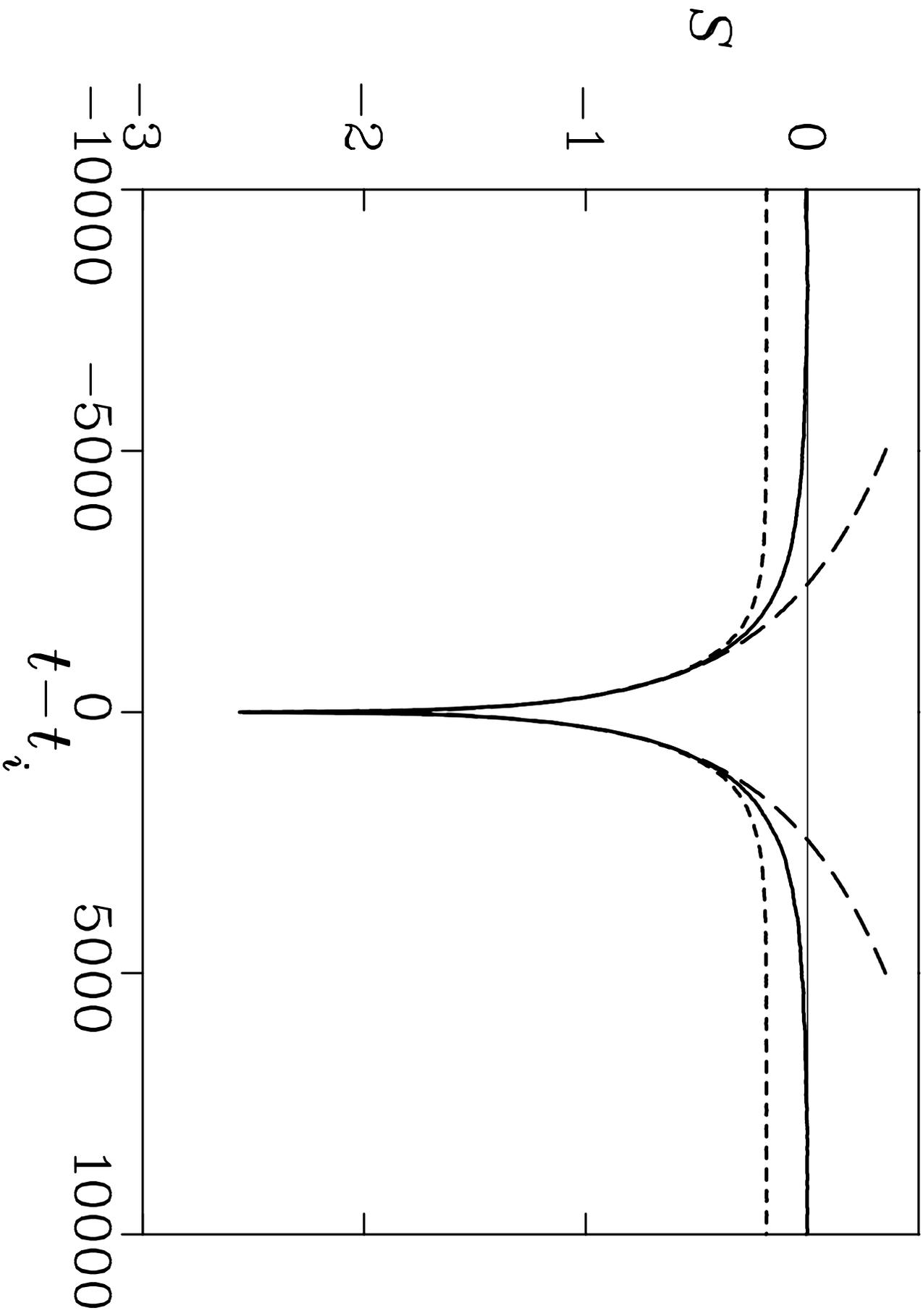
\end{document}